\documentclass[letterpaper]{article}
\usepackage{aaai}
\usepackage{times}
\usepackage{helvet}
\usepackage{courier}
\usepackage{amsmath}

\newtheorem{theorem}{Theorem}
\newtheorem{corollary}{Corollary}[theorem]

\newcommand{\systemname}[1]{\textit{#1}}

\newcommand{\alphabet}{\mathcal{A}}

\newcommand{\assignment}{\mathbf{A}}
\newcommand{\cassignment}{A}

\newcommand{\tass}[1]{\mathbf{T}#1}
\newcommand{\fass}[1]{\mathbf{F}#1}
\newcommand{\Tass}{\assignment^\mathbf{T}}
\newcommand{\Fass}{\assignment^\mathbf{F}}
\newcommand{\atom}[1]{atom(#1)}
\newcommand{\head}[1]{head(#1)}
\newcommand{\body}[1]{body(#1)}
\newcommand{\dneg}{not\ }
\newcommand{\domain}[1]{dom(#1)}
\newcommand{\range}[1]{range(#1)}
\newcommand{\scope}[1]{scope(#1)}

\newcommand{\encsup}{$S$}
\newcommand{\encbou}{$B$}
\newcommand{\encran}{$R$}
\newcommand{\encbouh}[1]{\encbou$_{#1}$}
\newcommand{\encranh}[1]{\encran$_{#1}$}

\newcommand{\citeay}[1]{\citeauthor{#1} \citeyear{#1}}
\newcommand{\citeap}[1]{\citeauthor{#1} (\citeyear{#1})}

\title{Reformulation of Global Constraints in Answer Set Programming}
\author{
Christian Drescher \\
Vienna University of Technology \\
Austria
\And
Toby Walsh \\
NICTA and University of New South Wales \\
Australia
}

\begin{document}
\maketitle

\begin{abstract}
We show that global constraints on finite domains like \emph{all-different} can be reformulated into answer set programs on which we achieve arc, bound or range consistency.
These reformulations offer a number of other 
advantages beyond providing the power of global
propagators to answer set programming. For example,
they provide other constraints with access to the state of the propagator by sharing variables. Such sharing can be used to improve propagation between constraints. Experiments with these encodings demonstrate their promise.
\end{abstract}

\section{Introduction}

There are several approaches to representing and solving constraint satisfaction problems: constraint programming (CP; \citeay{dechter03}, \citeay{robewa06a}), answer set programming (ASP; \citeay{baral03}), propositional satisfiability checking (SAT; \citeay{bihemawa09a}), its extension to satisfiability modulo theories (SMT; \citeay{niolti06a}), and many more. Each has its particular strengths: for example, CP systems support global constraints, ASP systems permit recursive definitions and offer default negation, whilst SAT solvers often exploit very efficient implementations. In many applications it would often be helpful to exploit the strengths of multiple approaches. Consider the problem of timetabling at a university~\cite{jaoijani09a}.  To model the problem, we need to express the mutual exclusion of events (for instance, 
we cannot place two events in the same room at the same time). A
straightforward representation of such constraint with clauses and rules uses
quadratic space. In contrast, global constraints such as \emph{all-different} typically supported by CP systems can give a much more concise encoding. On the other hand, there are features which are hard to describe in traditional constraint programming, like the temporary unavailability of a particular room. However, this is easy to represent with non-monotonic rules such as those used in ASP. Such rules also provide a flexible mechanism for defining new relations on the basis of existing ones.

Answer set programming has been put forward as a powerful paradigm to solve constraint  satisfaction problems. \citeap{niemela99a} shows that ASP embeds SAT but provides a more expressive framework from a knowledge representation point of view. Moreover, modern ASP solvers compete\footnote{\texttt{http://www.satcompetition.org}} with the best SAT solvers.
An empirical comparison of the performance of ASP and constraint logic programming (CLP; \citeay{jama94a}) systems on solving combinatorial problems conducted by \citeauthor{dofopo05a} shows ASP encodings to be more compact, more declarative, and highly competitive.
However, as some problems are more naturally modelled by using non-pro\-po\-si\-tional constructs, like resources or functions over finite domains, and by using global
constraints in particular, there is an increasing desire to handle constraints beyond pure ASP.

One approach to combining ASP and CP is to integrate theory-specific predicates into propositional formulas (motivated by SMT), and to extend the ASP solver's decision engine with a higher level proof procedure \cite{baboge05a,melgel08a,geossc09a}. However, the resulting systems have a number of limitations. First, they are tied to particular ASP and CP solvers. Second, the support for global constraints is limited. Third, communication between the ASP and CP solver is restricted.
Alternative techniques, such as reformulating constraints into ASP have received little attention.
The key contribution of our work is an investigation of reformulation in the context of answer set programming, illustrated by reformulations of the popular \emph{all-different} constraint. The resulting approach has been implemented in the new preprocessor \systemname{inca}. Empirical evaluation demonstrates its computational potential.


\section{Background}
\paragraph{Answer Set Programming}
A \emph{(normal) logic program} $\Pi$ over a set of primitive propositions $\alphabet$ is a finite set of rules of the form
$
a_0 \leftarrow a_1 , \dots , a_m, \dneg a_{m+1} , \dots , \dneg a_n 
$
where $0 \leq m \leq n$ and $a_i \in \alphabet$ are \emph{atoms} for $0 \leq i \leq n$. A \emph{literal} $\hat{a}$ is an atom $a$ or its default negation $\dneg a$.
For a rule~$r$, let $\head{r} = a_0$ be the \emph{head} of $r$ and $\body{r} = \{a_1 , \dots , a_m, \dneg a_{m+1} , \dots , \dneg a_n \}$ the \emph{body} of $r$. The set of atoms occurring in a logic program $\Pi$ is denoted by $\atom{\Pi}$, and the set of bodies in $\Pi$ is $\body{\Pi} = \{ \body{r} \mid r \in \Pi \}$. For regrouping bodies sharing the same head~$a$, define $\body{a} = \{ \body{r} \mid r \in \Pi,\ \head{r} = a \}$.
The semantics of a logic program is given by its answer sets, being total well-founded models of $\Pi$. For a formal introduction to ASP, we refer the reader to \citeap{baral03}.
The semantics of important extensions to logic programs, such as choice rules, integrity, and cardinality constraints, is given through program transformations that introduce additional propositions (cf. \citeay{siniso02a}).
A \emph{choice rule} allows for the non-deterministic choice over atoms in $\{a_1, \dots, a_k\}$ and has the form
$
\{a_0, \dots, a_k\} \leftarrow a_{k+1} , \dots , a_m, \dneg a_{m+1} , \dots , \dneg a_n.
$
An \emph{integrity constraint} of the form
$
\leftarrow a_1 , \dots , a_m, \dneg a_{m+1} , \dots , \dneg a_n 
$
is a short hand for a rule with an unsatisfiable head, and thus forbids its body to be satisfied in any answer set.
A~\emph{cardinality constraint} of the form
$
\leftarrow k \{\hat{a}_1 , \dots, \hat{a}_n\}
$
is interpreted as no $k$ literals of the set $\{\hat{a}_1 , \dots, \hat{a}_n\}$ are included in an answer set.
\citeauthor{siniso02a} provide a transformation that needs just $\mathcal{O}(nk)$ rules, introducing atoms $l(\hat{a}_i,j)$ to represent the fact that at least $j$ of the literals with index $\geq i$, i.e. the literals in $\{ \hat{a}_i, \dots, \hat{a}_n \}$,
 are in a particular answer set candidate. Then, the cardinality constraint can be encoded by an integrity constraint $\leftarrow l(\hat{a}_1,k)$ and the three following rules, where $1 \leq i \leq n$ and $1 \leq j \leq k$:
\[
\begin{array}{l@{\qquad\quad}r}
l(\hat{a}_i,j) \leftarrow l(\hat{a}_{i+1},j) & l(\hat{a}_i,j+1) \leftarrow \hat{a}_i, l(\hat{a}_{i+1},j)\\
l(\hat{a}_i,1) \leftarrow \hat{a}_i &
\end{array}
\]

\paragraph{Nogoods of Logic Programs}
We want to view inferences in ASP as unit-propagation on nogoods. 
Following \citeap{gekanesc07a}, inferences in ASP rely on atoms and program rules, which can be expressed by using atoms and bodies. Thus, for a program~$\Pi$, the \emph{domain} of Boolean assignments~$\assignment$ is fixed to $\domain{\assignment} = \atom{\Pi} \cup \body{\Pi}$.

Formally, a Boolean \emph{assignment} $\assignment$ is a set $\{ \sigma_1, \dots, \sigma_n \}$ of \emph{signed literals}~$\sigma_i$ for $1 \leq i \leq n$ of the form $\tass{a}$ or $\fass{a}$ where $a \in \domain{\assignment}$. $\tass{a}$ expresses that $a$ is assigned \emph{true} and $\fass{a}$ that it is \emph{false} in $\assignment$. (We omit the attribute \emph{Boolean} for assignments whenever clear from the context.) The complement of a signed literal~$\sigma$ is denoted by $\overline{\sigma}$, that is $\overline{\tass{a}} = \fass{a}$ and $\overline{\fass{a}} = \tass{a}$.
In the context of ASP, a \emph{nogood} is a set $\delta = \{ \sigma_1, \dots, \sigma_n \}$ of signed literals, expressing a constraint violated by any assignment~$\assignment$ such that $\delta \subseteq \assignment$.
For a nogood $\delta$, a signed literal $\sigma \in \delta$, and an assignment $\assignment$, we say that $\delta$ is \emph{unit} and $\overline{\sigma}$ is \emph{unit-resulting} if $\delta \setminus \assignment = \{\sigma\}$.
Let $\Tass = \{ a \in \domain{\assignment} \mid \tass{a} \in A \}$ the set of true propositions and $\Fass = \{ a \in \domain{\assignment} \mid \fass{a} \in A \}$ the set of false propositions. A \emph{total} assignment, that is $\Tass \cup \Fass = \domain{\assignment}$ and $\Fass \cup \Tass = \emptyset$, is a \emph{solution} for a set $\Delta$ of nogoods if $\delta \not\subseteq \assignment$ for all $\delta \in \Delta$.

As shown in \citeap{lee05a}, the answer sets of a logic program $\Pi$ correspond to the models of the completion of $\Pi$ that satisfy the loop formulas of all non-empty subsets of $\atom{\Pi}$. For $\beta = \{ a_1 , \dots , a_m, \dneg a_{m+1} , \dots , \dneg a_n \} \in \body{\Pi}$, define
\[
\Delta_\beta = \left\{ \begin{array}{l}
\{\tass{a_1}, \dots, \tass{a_m}, \fass{a_{m+1}}, \dots \fass{a_n}, \fass{\beta} \}, \\
\{\fass{a_1}, \tass{\beta}\}, \dots, \{\fass{a_m}, \tass{\beta}\}, \\
\{\tass{a_{m+1}}, \tass{\beta}\}, \dots, \{\tass{a_n}, \tass{\beta}\} \\
\end{array} \right\}.
\]
Intuitively, the nogoods in $\Delta_\beta$ enforce the truth of body~$\beta$ iff all its literals are satisfied.
For an atom $a \in \atom{\Pi}$ with $\body{a} = \{\beta_1, \dots, \beta_k\}$, let
\[
\Delta_a = \left\{ \begin{array}{l}
\{\fass{\beta_1}, \dots, \fass{\beta_k}, \tass{a} \}, \\
\{\tass{\beta_1}, \fass{a}\}, \dots, \{\tass{\beta_k}, \fass{a}\}
\end{array} \right\}.
\]
Then, the solutions for $\Delta_\Pi = \bigcup_{\beta \in \body{\Pi}} \Delta_\beta \cup \bigcup_{a \in \atom{\Pi}} \Delta_a$ correspond to the models of the completion of $\Pi$. Loop formulas, expressed in the set of nogoods~$\Lambda_\Pi$, have to be added to establish full correspondence to the answer sets of $\Pi$.
Typically, solutions for $\Delta_\Pi \cup \Lambda_\Pi$ are computed by applying \emph{conflict-driven nogood learning} (CDNL; \citeap{gekanesc07a}). This combines search and propagation by recursively assigning the value of a proposition and using \emph{unit-propagation} to determine logical consequences of an assignment \cite{mitchell05a}.

\paragraph{Constraint Satisfaction Problem}
The classic definition of a constraint satisfaction problem is as follows (cf. \citeay{robewa06a}). A \emph{constraint satisfaction problem} is a triple $(V,D,C)$ where $V$ is a set of \emph{variables} $V = \{v_1, \dots , v_n\}$, $D$ is a set of finite \emph{domains} $D=\{D_1, \dots , D_n\}$ such that each variable~$v_i$ has an associated domain $\domain{v_i} = D_i$, and $C$ is a set of \emph{constraints}. A constraint~$c$ is a pair~$(R_S,S)$ where $R_S$ is a $k$-ary \emph{relation} on the variables in $S \subseteq V^k$, called the \emph{scope} of $c$. In other words, $R_S$ is a subset of the Cartesian product of the domains of the variables in $S$. To access the relation and the scope of $c$ define $\range{c} = R_S$ and $\scope{c} = S$. For a \emph{(constraint variable) assignment} $\cassignment : V \to \bigcup_{v \in V} dom(v)$ and a constraint $c = (R_S, S)$ with $S = (v_1, \dots, v_k)$, define $\cassignment(S) = (\cassignment(v_1), \dots, \cassignment(v_k))$, and call $c$ \emph{satisfied} if $\cassignment(S) \in \range{c}$. Given this, define the set of constraints satisfied by $\cassignment$ as
$
sat_C(\cassignment) = \{ c \mid \cassignment(\scope{c}) \in \range{c},\ c \in C\}.
$

A binary constraint~$c$ has $|scope(c)|=2$. For example,
$v_1 \neq v_2$ ensures that $v_1$ and $v_2$ take different
values. A global (or $n$-ary) constraint~$c$ has parametrized
scope. For example, 
the \emph{all-different} constraint ensures that
a set of variables, $\{v_1,\ldots,v_n\}$ take all different values. 
This can be decomposed into $O(n^2)$ binary
constraints, $v_i \neq v_j$ for $i<j$. However, as 
we shall see, such reformulation can hinder inference. 
An assignment~$\cassignment$ is a \emph{solution} iff it satisfies all constraints in $C$.

Constraint solvers typically use backtracking search to explore the space of partial assignments. Various heuristics affecting, for instance, the variable selection criteria and the ordering of the attempted values, can be used to guide the search. Each time a variable is assigned a value, a deterministic propagation stage is executed, pruning the set of values to be attempted for the other variables, i.e., enforcing a certain type of local consistency.

A binary constraint~$c$ is called \emph{arc consistent} iff when a variable~$v_1 \in \scope{c}$ is assigned any value~$d_1 \in \domain{v_1}$, there exists a consistent value~$d_2 \in \domain{v_2}$ for the other variable~$v_2$.
An $n$-ary constraint~$c$ is \emph{hyper-arc consistent} or \emph{domain consistent} iff when a variable~$v_i \in \scope{c}$ is assigned any value~$d_i \in \domain{v_i}$, there exist compatible values in the domains of all the other variables~$d_j \in \domain{v_j}$ for all $1 \leq j \leq n,\ j \neq i$ such that $(d_1, \dots, d_n) \in \range{c}$.

Relational consistency \cite{debe97a} extends the concept of local consistency. I.e. a constraint~$c$ is \emph{relationally $k$-arc consistent} if any consistent assignment of a $k$-elementary subset of variables from $\scope{c}$ extends to a consistent assignment of all variables in $\scope{c}$.

The concepts of bound and range consistency are defined for constraints on ordered intervals.
Let $min(D_i)$ and $max(D_i)$ be the minimum value and maximum value of the domain~$D_i$. A constraint~$c$ is \emph{bound consistent} iff when a variable~$v_i$ is assigned $d_i \in \{min(\domain{v_i}), max(\domain{v_i})\}$ (i.e. the minimum or maximum value in its domain), there exist compatible values between the minimum and maximum domain value for all the other variables in the scope of the constraint. Such an assignment is called a \emph{bound support}. A constraint is \emph{range consistent} iff when a variable is assigned any value in its domain, there exists a bound support. Notice that range consistency is in between domain and bound consistency, where domain consistency is the strongest of the three formalisms.

\section{Encoding Global Constraints in ASP}
In this section we explain how to reformulate multi-valued variables and constraints on finite domains into a logic program under answer set semantics. In what follows, we assume $\domain{v} = \lbrack 1, d\rbrack$ for all $v \in V$ to save the reader from multiple superscripts.

\paragraph{Direct Encoding}
A popular choice is called the \emph{direct encoding} \cite{wa00}. In the direct encoding, a propositional variable $e(v, i)$, representing $v = i$, is introduced for each value~$i$ that can be assigned to the constraint variable~$v$. Intuitively, the proposition $e(v, i)$ is true if $v$ takes the value $i$, and false if $v$ takes a value different from $i$. For each $v$, the truth-assignments of atoms $e(v, i)$ are encoded by a choice rule (1). Furthermore, there is an integrity constraint (2) to ensure that $v$ takes at least one value, and a cardinality constraint (3) that ensures that $v$ takes at most one value.
\begin{align}
\{ e(v, 1), \dots, e(v, d) \} &\leftarrow \\
&\leftarrow \dneg e(v, 1), \dots, \dneg e(v, d) \\
&\leftarrow 2\ \{ e(v, 1), \dots, e(v, d) \}
\end{align}
In the direct encoding, each forbidden combination of values in a constraint is expressed by an integrity constraint. On the other hand, when a relation is represented by allowed combinations of values, all forbidden combinations have to be deduced and translated to integrity constraints. Unfortunately, the direct encoding of constraints hinders propagation:
\begin{theorem}
Enforcing arc consistency on the binary decomposition of the original constraint prunes more values from the variables domain than unit-propagation on its direct encoding.
\end{theorem}

\paragraph{Support Encoding}
The \emph{support encoding} has been proposed to
tackle this weakness \cite{gent02}. A \emph{support} for a constraint variable~$v$ to take the value~$i$ across a constraint~$c$ is the set of values $\{i_1, \dots, i_m\} \subseteq \domain{v'}$ of another variable in~$v' \in \scope{c}\setminus \{v\}$ which allow $v = i$, and can be encoded as follows, extending (1--3):
\[
\leftarrow e(v, i), \dneg e(v', i_1), \dots, \dneg e(v', i_m)
\]
This integrity constraint can be read as whenever $v = i$, then at least one of its supports must hold.
In the support encoding, for each constraint~$c$ there is one support for each pair of distinct variables $v, v' \in \scope{c}$, and for each value~$i$.
\begin{theorem}
Unit-propagation on the support encoding enforces arc consistency on the binary decomposition of the original constraint.
\end{theorem}
We illustrate this approach on an encoding of the global \emph{all-different} constraint. For variables $v, v'$ and value $i$ it can be reduced from the definition by using the equivalence covered by (2--3) to
\[
\leftarrow e(v, i), e(v', i).
\]
Observe, that this is also
the direct encoding of the binary decomposition of the global \emph{all-different} constraint. However, this observation does not hold in general for all constraints.
As discussed in the Background section of this paper, we can express above condition as $\mathcal{O}(d)$ cardinality constraints:
\begin{align}
&\leftarrow 2\ \{ e(v_1, i), \dots, e(v_n, i) \}
\end{align}

\begin{corollary}
Unit-propagation on (1--4) enforces arc consistency on the binary decomposition of the global \emph{all-different} constraint in $\mathcal{O}(nd^2)$ down any branch of the search tree.
\end{corollary}

\paragraph{k-support Encoding}
The support encoding can be generalized to the \emph{$k$-support encoding} \cite{behewa03a} representing supports on subsets of $\scope{c}$ for an assignment of another $k$-elementary subset of variables in $\scope{c}$. More formal, a $k$-support~$S$ for an assignment $\cassignment$ of $k$~variables from $\scope{c}$, say $v_1 = i_1, \dots, v_k = i_k$, is an assignment $v'_1 = i'_1, \dots, v'_l = i'_l$ such that $\{v'_1, \dots, v'_l\} \subseteq \scope{c} \setminus \{v_1, \dots, v_k\}$ which allows $\cassignment$. We introduce a \emph{support-variable}~$s$, that evaluates to true iff $S$ holds:
\[
s \leftarrow e(v'_1, i_1), \dots, e(v'_l, i'_l)
\]
Furthermore, let $\{S_1, \dots, S_m\}$ be the set of all $k$-supports of $\cassignment$. A $k$-support rule for $\cassignment$ is defined as
\[
\leftarrow e(v_1, d_1), \dots, e(v_k, d_k), \dneg s_1, \dots, \dneg s_m
\]
meaning that as long as $\cassignment$ holds then at least one of its $k$-supports $S_1, \dots S_m$ must hold.
In the $k$-support encoding, for each constraint~$c$ there is one $k$-support rule for each assignment~$\cassignment$ of $k$~variables from $\scope{c}$.
\begin{theorem}
Unit-propagation on the $k$-support encoding enforces relational $k$-arc consistency on the original constraint.
\end{theorem}

\paragraph{Range Encoding}
In the \emph{range encoding}, a propositional variable~$r(v, l, u)$ is introduced for all $\lbrack l, u \rbrack \subseteq \lbrack 1, d \rbrack$ to represent whether the value of~$v$ is between $l$ and $u$. For each range~$\lbrack l, u \rbrack$, the following $\mathcal{O}(nd^2)$ rules encode $v \in \lbrack l , u \rbrack$ whenever it is safe to assume that $v \not\in \lbrack 1, l-1 \rbrack$ and $v \not\in \lbrack u+1, d\rbrack$, and enforce a consistent set of ranges such that $v \in \lbrack l, u\rbrack \Rightarrow v \in \lbrack l-1, u\rbrack \land v \in \lbrack l, u+1\rbrack$:
\begin{align}
r(v, l, u) &\leftarrow \dneg r(v, 1, l-1), \dneg r(v, u+1, d) \\
&\leftarrow r(v, l-1, u), \dneg r(v, l, u) \\
&\leftarrow r(v, l, u+1), \dneg r(v, l, u)
\end{align}
Constraints are encoded into integrity constraints representing conflict regions. When the combination $v_1 \in \lbrack l_1, u_1\rbrack, \dots, v_n \in \lbrack l_n, u_n\rbrack$ violates the constraint, the following rule is added:
\[
\leftarrow r(v_1, l_1, u_1), \dots, r(v_n, l_n, u_n)
\]
\begin{theorem}
Unit-propagation on the range encoding enforces range consistency on the original constraint.
\end{theorem}
A propagator for the global \emph{all-different} constraint that enforces range consistency pruning Hall intervals has been proposed by \citeap{le96a} and encoded to SAT by \citeap{bekanaquwa09a}.
An interval~$\lbrack l, u \rbrack$ is a \emph{Hall interval} iff $|\{ v \mid dom(v) \subseteq \lbrack l, u \rbrack \}| = u - l + 1$. In other words, a Hall interval of size~$k$ completely contains the domains of $k$~variables. Observe that in any bound support, the variables whose domains are contained in the Hall interval consume all values within the Hall interval, whilst any other variable must find their support outside the Hall interval.
The following reformulation of the global \emph{all-different} constraint will permit us to achieve range consistency via unit propagation. It ensures that no interval $\lbrack l, u\rbrack$ can contain more variables than its size. 
\begin{align}
&\leftarrow u-l+2\ \{ r(v_1, l, u), \dots, r(v_n, l, u) \}
\end{align}
This simple reformulation can simulate a complex propagation algorithm like \citeauthor{le96a}'s with a similar overall complexity of reasoning.
\begin{corollary}
Unit-propagation on (5--8) enforces range consistency on the global \emph{all-different} constraint in $\mathcal{O}(nd^3)$ down any branch of the search tree.
\end{corollary}
A hybrid that links the range encoding of $v$ to its direct representation extends the range encoding as follows, for each $i \in \domain{v}$:
\[
\begin{array}{r@{\ \leftarrow\ }l}
e(v, i) & r(v, i, i)\\
& e(v, i), \dneg r(v, i, i)\\
\end{array}
\]
These rules encode the equivalence $v=i \Leftrightarrow v \in \lbrack i, i \rbrack$.

\paragraph{Bound Encoding}
A last encoding is called the \emph{bound encoding} \cite{crba94a}. In the bound encoding, a propositional variable~$b(v, i)$ is introduced for each value $i$ to represent that the value of~$v$ is bounded by~$i$. That is, $v \leq i$ if $b(v,i)$ is assigned \emph{true}, and $v > i$ if $b(v,i)$ is assigned \emph{false}. Similar to the direct encoding, for each $v$, the truth-assignments of atoms~$b(v, i)$ are encoded by a choice rule (9). In order to ensure that assignments represent a consistent set of bounds, the condition $v \leq i \Rightarrow v \leq i+1$ is posted as integrity constraints (10) $\forall i \in \lbrack 1, d-1 \rbrack$. Another integrity constraint (11) encodes $v \leq d$, that at least one value must be assigned to $v$:
\begin{align}
\{ b(v, 1), \dots, b(v, d) \} &\leftarrow \\
&\leftarrow  b(v, i), \dneg b(v, i+1) \\
&\leftarrow \dneg b(v, d)
\end{align}
Constraints are encoded into integrity constraints representing conflict regions similar to the range encoding. When all combinations in the region
\[
l_1 < v_1 \leq u_1, \dots, l_n < v_n \leq u_n
\]
violate a constraint, the following rule is added:
\[
\leftarrow b(v_1, u_1), \dots, b(v_n, u_n), \dneg b(v_1, l_1), \dots, \dneg b(v_n, l_n)
\]
\begin{theorem}
Unit-propagation on the bound encoding enforces bound consistency on the original constraint.
\end{theorem}
In order to get a representation of the global \emph{all-different} constraint that can only prune bounds, the bound encoding for variables is linked to (8) as follows:
\begin{align}
r(v, l, u) &\leftarrow \dneg b(v, l-1), b(v, u) \\
&\leftarrow r(v, l, u), b(v, l-1) \\
&\leftarrow r(v, l, u), \dneg b(v, u)
\end{align}
\begin{corollary}
Unit-propagation on (8--14) enforces bound consistency on the global \emph{all-different} constraint in $\mathcal{O}(nd^2)$ down any branch of the search tree.
\end{corollary}
Note that an upper bound $h$ can be posted on the size of Hall intervals. The resulting encoding with only those cardinality constraints (5) for which $u - l + 1 \leq h$ detects Hall intervals of size at most $h$, and therefore enforces a weaker level of consistency.

To access the value of $v$, the bound encoding can be extended to a hybrid by adding the following rules to the bound encoding for each $i \in \lbrack 1, d \rbrack$:
\[
\begin{array}{r@{\ \leftarrow\ }l}
e(v, i) & b(v, i), \dneg b(v, i-1) \\
& e(v, i), \dneg b(v, i)\\
& e(v, i), b(v, i-1)
\end{array}
\]
The first rule enforces $e(v, i)$ to be true if possible values for $v$ are bound to the singleton $i$, i.e. $v \leq i$ and $v \not\leq i-1$ are in the assignment. On the other hand, the condition $v = i \Rightarrow v \leq i \land v \not\leq i-1$ is represented as integrity constraints.

\paragraph{Non-ground Logic Programs}
Although our semantics is propositional, atoms in $\alphabet$ and can be constructed from a first-order signature $\Sigma = (\mathcal{F}, \mathcal{V}, \mathcal{P})$, where
$\mathcal{F}$ is a set of function symbols (including constant symbols),
$\mathcal{V}$ is a denumerable collection of first-order variables, and
$\mathcal{P}$ is a set of predicate symbols.
The logic program over $\alphabet$ is then obtained by a grounding process, systematically substituting all occurrences of variables $\mathcal{V}$ by terms in $\mathcal{T}(\mathcal{F})$, where $\mathcal{T}(\mathcal{F})$ denotes the set of all ground terms over $\mathcal{F}$. Atoms in $\alphabet$ are formed from predicate symbols $\mathcal{P}$ and terms in $\mathcal{T}(\mathcal{F})$.

\section{Experiments}
To evaluate these reformulations, we conducted experiments on encodings containing \emph{all-different} and \emph{permutation} constraints.
The global \emph{permutation} constraint is a special case of \emph{all-different} when the number of variables is equal to the number of all their possible values. A reformulation of \emph{permutation} extends (4) by
\[
\leftarrow \dneg e(v_1, i), \dots, \dneg e(v_n, i)
\]
or (8) by the following rule where $1 \leq l \leq u \leq k$:
\[
\leftarrow d-u+l\ \{ \dneg r(v_1, l, u), \dots, \dneg r(v_n, l, u) \}
\]
This can increase propagation.
Our reformulations have been implemented within the prototypical preprocessor \systemname{inca}\footnote{\texttt{http://potassco.sourceforge.net/} provides the systems \systemname{clasp}, \systemname{clingo}, \systemname{gringo}, \systemname{inca}, and the benchmark set} which compiles an (extended) logic programs with high-level statements for global constraints, constraint variables, first-order variables, function symbols, and aggregates, etc. in linear time and space, such that the logic program can be obtained by a \emph{grounding} process.
Experiments consider \systemname{inca} in different settings using different reformulations. We denote the support encoding of the global constraints by \encsup, the bound encoding of the global constraints by \encbou, and the range encoding of the global constraints by \encran. To explore the impact of small Hall intervals, we also tried \encbouh{k} and \encranh{k}, an encoding of the global constraints with only those cardinality constraints (8) for which $u-l+1 \leq k$. The consistency achieved by \encbouh{k} and \encranh{k} is therefore weaker than full bound and range consistency, respectively.

We also include the pure CP system \systemname{gecode}\footnote{\texttt{http://www.gecode.org/}} (3.2.0), and the integrated system \systemname{ezcsp}\footnote{\texttt{http://krlab.cs.ttu.edu/\~{}marcy/ezcsp/}} (1.6.9; \citeay{ba09a}) in our empirical analysis. The latter combines the grounder \systemname{gringo} (2.0.3) and ASP solver \systemname{clasp} (1.3.0) with \systemname{sicstus}\footnote{\texttt{http://www.sics.se/sicstus/}} (4.0.8) as a constraint solver.
Since \systemname{inca} is a pure preprocessor, we select the ASP system \systemname{clingo} (2.0.3) as its backend to provide a representative comparison with \systemname{ezcsp}. Note that \systemname{clingo} stands for \systemname{clasp} on \systemname{gringo} and combines both systems in a monolithic way.
All experiments were run on a 2.00~GHz PC under Linux. We report results in seconds, where each run was limited to 600 s time and 1 GB RAM.

\paragraph{Pigeon Hole Problem}
The \emph{pigeon hole problem} (PHP) is to show that it is impossible to put $n$ pigeons into $n-1$ holes if each pigeon must be put into a distinct hole.
Clearly, our bound and range reformulations are faster compared to weaker encodings (see Table \ref{tab:php}). It appears that \systemname{sicstus}' and \systemname{gecode}'s default configuration uses filtering algorithms for the global \emph{all-different} constraint achieve arc consistency on its binary decomposition. However, on such problems, detecting large Hall intervals is essential.
\begin{table}
\centering
\begin{tabular}{cccccccc} \hline\hline
$n$ & \encsup & \encbouh{3} & \encbou & \encranh{3} & \encran & \systemname{ezcsp} & \systemname{gecode} \\ \hline
10 & 5 & \textbf{$<$1} & \textbf{$<$1} & $<$1 & \textbf{$<$1} & 2 & \textbf{$<$1} \\
11 & 46 & 1 & \textbf{$<$1} & 2 & \textbf{$<$1} & 17 & 9 \\
12 & 105 & 4 & \textbf{$<$1} & 3 & \textbf{$<$1} & 184 & 104 \\
13 & --- & 25 & \textbf{$<$1} & 30 & \textbf{$<$1} & --- & --- \\
14 & --- & 125 & \textbf{$<$1} & 197 & \textbf{$<$1} & --- & --- \\
15 & --- & --- & \textbf{$<$1} & --- & \textbf{$<$1} & --- & --- \\
16 & --- & --- & \textbf{$<$1} & --- & \textbf{$<$1} & --- & --- \\ \hline\hline
\end{tabular}
\caption{Runtime results in seconds for PHP. \label{tab:php}}
\end{table}

\paragraph{Latin Squares}
A \emph{Latin square} is an $n \times n$-table filled with $n$ different elements such that each element occurs exactly once in each row and each column of the table. The \emph{Latin square puzzle}~(LSP) is to determine whether a partially filled table can be completed in such a way that a Latin square is obtained.
Randomly generated LSP has been proposed as a benchmark domain for CP systems by \citeauthor{gose97a} since it combines the features of purely random problems and highly structured problems.
\begin{table}
\centering
\begin{tabular}{cccccccc} \hline\hline
\% & \encsup & \encbouh{3} & \encbou & \encran & \systemname{ezcsp} & \systemname{gecode} & \systemname{gecode}$_{B}$ \\ \hline
10 & \textbf{3} & 5 & 8 & 7 &30 (7)  & 2 (4) & $<$1 (1) \\
20 & \textbf{2} & 5 & 8 & 7 &21 (20) & 5 (4) & $<$1 (3) \\
30 & \textbf{2} & 5 & 8 & 7 &10 (30) & 3 (13) & 1 (5) \\
35 & \textbf{2} & 5 & 8 & 7 &22 (24) &14 (13) & 6 (7) \\
40 & \textbf{2} & 5 & 8 & 7 &52 (29) &12 (20) & 6 (9) \\
45 & \textbf{2} & 5 & 8 & 7 &36 (35) &18 (25) & 6 (13) \\
50 & \textbf{2} & 5 & 8 & 7 &36 (50) &25 (32) & 6 (18) \\
55 & \textbf{2} & 4 & 8 & 7 &61 (51) &20 (41) &31 (29) \\
60 & \textbf{2} & 4 & 8 & 7 &60 (63) &36 (51) &27 (35) \\
70 & \textbf{2} & 4 & 7 & 6 &70 (66) &28 (45) &17 (27) \\
80 & \textbf{2} & 4 & 7 & 5 &16 (18) &17 (13) & 7 (7) \\
90 & 2 & 4 & 7 & 5 & \textbf{1} & $<$1 (1) & 3 \\ \hline\hline
\end{tabular}
\caption{Average times over 100 runs on LSP. Timeouts are given in parenthesis, if any.\label{tab:qcp}}
\vspace{-1\baselineskip}
\end{table}
Table \ref{tab:qcp} compares the runtime for solving LSP problems of size $20 \times 20$ where the first column gives the percentage of preassigned values. We included \systemname{gecode} with algorithms that enforce bound and domain consistency, denoted as \systemname{gecode}$_{B}$ and \systemname{gecode}$_{D}$ (not shown due to space constraints), in the experiments. Our analysis exhibits phase transition behaviour of the systems \systemname{ezcsp}, \systemname{gecode}, and \systemname{gecode}$_{B}$, while our Boolean encodings and \systemname{gecode}$_{D}$ solve all problems within seconds. Interestingly, learning constraint interdependencies as in our approach is sufficient to tackle LSP. In fact, most of the time for $S$, $B_k$, $R_k$ is spent on grounding, but not for solving the actual problem.

\paragraph{Graceful Graphs}
A labelling $f$ of the nodes of a graph $(V,E)$ is \emph{graceful} if $f$ assigns a unique label~$f(v)$ from $\{0,1,\dots,|E|\}$ to each node $v \in V$ such that, when each edge $(v,w) \in E$ is assigned the label $|f(v)-f(w)|$, the resulting edge labels are distinct. The problem of determining the existence of a graceful labelling of a graph (GGP) has been modelled in CP \citeap{pesm03a}, using auxillary variables $d(v,w)$ for edge labels. We represent the equivalence $d(v,w) = |f(v)-f(w)|$ in the direct encoding which weakens the overall consistency. Our experiments consider double-wheel graphs $DW_n$ composed by two copies of a cycle with $n$ vertices, each connected to a central hub.
\begin{table}
\centering
\begin{tabular}{cccccccc} \hline\hline
$n$ & \encsup & \encbouh{1} & \encbouh{3} & \encbou & \encran & \systemname{ezcsp} & \systemname{gecode} \\ \hline
$3$ & 11 &  4 &  6 &  9 &  6 & 6 & \textbf{2} \\
$4$ &  1 &  2 &  1 &  3 &  3 & \textbf{$<$1} & \textbf{$<$1} \\
$5$ &  4 &  5 &  4 & 13 & 12 & 1 & \textbf{$<$1} \\
$6$ &  7 & 11 & 18 & 48 & 21 & \textbf{1} & 7 \\
$7$ & 24 & 28 & 68 &228 & 60 & \textbf{18} & --- \\
$8$ & 48 & 68 & ---  &208 & 58 & \textbf{4} & --- \\
$9$ & \textbf{83} &106 &200 &487 & ---  & 390 & --- \\ \hline\hline
\end{tabular}
\caption{Runtime results in seconds for GGP. \label{tab:ggp}}
\vspace{-1\baselineskip}
\end{table}
Table \ref{tab:ggp} shows that our encodings compete with \systemname{ezcsp} and outperform \systemname{gecode}, where the support encoding performs better than bound and range encodings. In most cases, the branching heuristic used in our approach appears to be misled by the extra variables introduced in $B_k$ and $R_k$. That explains some of the variability in the runtimes.

\section{Conclusions}

We have reformulated global and other constraints into answer set programs. In particular, we have investigated various generic ASP encodings for constraints on finite domains and proved which level of consistency unit-pro\-pa\-ga\-tion achieves on them.
Our techniques were formulated as preprocessing and can be applied to any ASP system without changing its source code, which allows for programmers to select the ASP solver that best fit their needs. We have empirically evaluated the performance of
such an approach on benchmarks from CP and found that such reformulations outperform integrated ASP(CP) systems as well as pure CP solvers.
%
%
Our future works includes the reformulation of other useful
global constraints into answer set programming like the
\emph{regular} constraint, as well as global constraints
like \emph{lex} which are very useful for symmetry breaking .

\end{document}